\documentclass{article}
\usepackage{psfig}
\begin{document}
  \title{Dynamic confinement of jets by magneto-torsional oscillations}
  \author{G.S. Bisnovatyi-Kogan
  \thanks{Space Research Institute, Moscow, Russia, Profsoyuznaya
  84/32, Moscow 117997, Russia;  gkogan@iki.rssi.ru}}
  \date{}
  \maketitle
  \begin{abstract}

Many quasars and active galactic nuclei (AGN) appear in radio,
 optical, and X-ray maps,
as a bright nuclear sources from which emerge single or double
long, thin jets  (Thomson et al., 1993). When observed with high angular resolution
these jets show structure with bright knots separated by
relatively dark regions. High percentages of polarization, sometimes
more then 50\% in some objects, indicates the nonthermal nature of the
radiation which
is well explained as the synchrotron radiation of the relativistic
electrons in an ordered magnetic field.

A strong collimation of jets is most probably connected with ordered
magnetic fields. The mechanism of magnetic collimation, first suggested by
Bisnovatyi-Kogan et al. (1969), was based on the initial charge separation,
leading to creation of oscillating electrical current, which produces azimuthal
magnetic field, preventing jet expansion and disappearance. Here we consider
magnetic collimation, connected with torsional oscillations of a cylinder with
elongated magnetic field. Instead of initial blobs with charge separation, we
consider a cylinder with a periodically distributed initial rotation
around the cylinder axis. The stabilizing azimuthal magnetic field is created here by
torsional oscillations, where charge separation is not necessary. Approximate simplified
model is developed. Ordinary differential equation is derived, and solved numerically,
what gives a possibility to estimate quantitatively the range of parameters where
jets may be stabilized by torsional oscillations.

\end{abstract}

\medskip


\section{Introduction}

Objects of different scale and nature in the universe: from young
and very old stars to galactic nuclei show existence of collimated outbursts
- jets. Geometrical sizes of jets
lay between parsecs and megaparsecs.
The origin of jets is not well understood and only several
qualitative mechanisms are proposed which are not
justified by calculations. Theory of jets must give answers to three
main questions: how jets are formed? how are they stabilized? how do they
radiate? The last question is related to the problem of the origin if
relativistic particles in outbursts from AGN, where synchrotron emission
is observed. Relativistic particles, ejected from the central machine
rapidly loose their energy so the problem arises of particle acceleration
inside the jet, see review of Bisnovatyi-Kogan (1993).

 It is convenient sometimes to investigate
jets in a simple model of infinitely long circular cylinder, Chandrasekhar \& Fermi (1953).
 The magnetic field in the collimated jets determines its
direction, and the axial current may stabilize the jet's elongated form at
large distances from the source (e.g. in AGNs), (Bisnovatyi-Kogan et al., 1969),
see also Istomin and Pariev (1996), Beskin (2005). When observed
with high angular resolution these jets show a structure with
bright knots separated by relatively dark regions (Thomson et al., 1993).
 High percentages of polarization, sometimes
exceeding 50\%, indicate the nonthermal nature of
the radiation, which is well explained as synchrotron emission of
the relativistic electrons in an ordered magnetic field.
Estimates of the lifetime of these electrons, based on the
observed luminosities and spectra, often give values much less
than the kinematic ages $t_k=d/c $, where $d$ is the distance of the
emitting point from the central source. There is a necessity of continuous
re-acceleration of the electrons in the jets in order to explain
the observations. The acceleration mechanism for electrons in
extragalactic jets proposed by Bisnovatyi-Kogan \& Lovelace (1995), considers that intense
long-wavelength electromagnetic oscillations accompany a
relativistic jet, and the electromagnetic
wave amplitudes envisioned are sufficient to give in situ
acceleration of electrons to the very high energies observed $>
10^{13} $ eV. It was assumed that jets are formed by a sequence of
outbursts from the nucleus with considerable charge separation at
the moment of the outburst. The direction of motion of
the outbursts is determined by the large-scale magnetic field.
When the emitted wave is strong enough it washes out the medium
around and the density can become very small, consisting only of the
accelerated particles. The action of the oscillating knot is
similar to the action of a pulsar, considered as an inclined magnetic rotator. Both emit
strong electromagnetic waves, which could effectively accelerate particles.
 The model of enhanced oscillations of the cylinder,
 and electromagnetic field around it, have been studied by Bisnovatyi-Kogan (2004).

 Here we consider stabilization of a jet by pure magnetohydrodynamic mechanism,
 connected with torsional oscillations. Such type of oscillations in neutron stars had been
 considered by Bastrukov et al. (2002).
 We suggest that the matter in the jet is
 rotating, and different parts of the jet rotate in different directions. Such distribution of
the rotational velocity produces azimuthal magnetic field, which prevents
a disruption of the jet. The jet remains to be in a dynamical equilibrium, when
it is representing a periodical, or quasiperiodical structure along the axis,
and its radius is oscillating with time all along the axis. The space and time
period of oscillations depend on the conditions at jet formation: the length scale, the
amplitude of the rotational velocity, and the strength of the magnetic field. The time
period of oscillations should be obtained during construction of the dynamical model,
what also should show at which input parameters may exist a long jet, stabilized by
torsional oscillations.

2D nonstationary MHD calculations are needed, to solve the problem numerically. Here we
construct a very simplified model of this phenomena, which, nonetheless, permits to confirm
the reality of such stabilization, to estimate the range of parameters at which is takes
place, and the connection between the time and space scales, magnetic field strength,
and the amplitude of rotational velocity.

 \section{Axially symmetric MHD equations  }

Axially symmetric MHD equations at $\frac{\partial}{\partial\phi}$, for the perfect gas with
an infinite conductivity, are written, in cylindric coordinates $(r,\phi,z)$, in the form
(Landau and Lifshits, 1982; Bisnovatyi-Kogan, 2001)

\begin{equation}
\label{eq1}
 \frac{\partial{v_r}}{\partial t}+v_r\,\frac{\partial{v_r}}{\partial r}
  +v_z\,\frac{\partial{v_r}}{\partial z}-{v_\varphi^2\over r}=
  -{1\over \rho}\,\frac{\partial P}{\partial r}-\frac{\partial{\varphi_G}}{\partial r}
 +{1\over \rho c}\left(j_\varphi B_z-j_z B_\varphi\right),
 \end{equation}

\begin{equation}
\label{eq2}
 \frac{\partial{v_\varphi}}{\partial t}+v_r\,\frac{\partial{v_\varphi}}{\partial r}
  +v_z\,\frac{\partial{v_\varphi}}{\partial z}+{v_r v_\varphi \over r}=
{1\over \rho c}\left(j_z B_r-j_r B_z\right),
 \end{equation}

\begin{equation}
\label{eq3}
 \frac{\partial{v_z}}{\partial t}+v_r\,\frac{\partial{v_z}}{\partial r}
  +v_z\,\frac{\partial{v_z}}{\partial z}=
  -{1\over \rho}\,\frac{\partial P}{\partial z}-\frac{\partial{\varphi_G}}{\partial z}
 +{1\over \rho c}\left(j_r B_\varphi -j_\varphi B_r\right),
 \end{equation}

\begin{equation}
\label{eq4}
  \frac{\partial{\rho}}{\partial t}+v_r\,\frac{\partial{\rho}}{\partial r}
  +v_z\,\frac{\partial{\rho}}{\partial z}+
  \rho \left[{1\over r}\,\frac{\partial{(rv_r)}}{\partial r}
  +\frac{\partial{v_z}}{\partial z}\right]=0,
 \end{equation}

\begin{equation}
\label{eq5}
  \frac{\partial{B_r}}{\partial t}=-{\partial\over \partial z}
  \left(v_zB_r-v_rB_z\right),
 \end{equation}

\begin{equation}
\label{eq6}
 \frac{\partial{B_\varphi}}{\partial t}={\partial\over \partial z}
  \left(v_\varphi B_z-v_z B_\varphi\right)-
  {\partial\over \partial r}\left(v_r B_\varphi-v_\varphi B_r\right),
 \end{equation}

\begin{equation}
\label{eq7}
  \frac{\partial{B_z}}{\partial t}={1\over r}\,{\partial\over \partial r}
  \left[r\left(v_zB_r-v_rB_z\right)\right],
 \end{equation}

\begin{equation}
\label{eq8}
  {1\over r}\,{\partial\over \partial r}(rB_r)+\frac{\partial{B_z}}{\partial z}=0,
 \end{equation}

\begin{equation}
\label{eq9}
  j_r=-{c\over 4\pi}\,\frac{\partial B_\varphi}{\partial z},
 \end{equation}

\begin{equation}
\label{eq10}
  j_\varphi={c\over 4\pi}\left(\frac{\partial{B_r}}{\partial z}
  -\frac{\partial{B_z}}{\partial r}\right),
 \end{equation}

\begin{equation}
\label{eq11}
  j_z={c\over 4\pi r}\,{\partial\over \partial r}(rB_\varphi),
 \end{equation}

\begin{equation}
\label{eq12}
  {1\over r}\,{\partial\over \partial r}\left(r\frac{\partial{\varphi_G}}{\partial r}\right)+
  {\partial^2\varphi_G\over \partial z^2}=4\pi G\rho,
 \end{equation}

\begin{equation}
\label{eq13}
  \frac{\partial E}{\partial t}+v_r\,\frac{\partial E}{\partial r}
  +v_z\,\frac{\partial E}{\partial z}+{P\over \rho}\,
  \left[{1\over r}\,\frac{{(rv_r)}}{\partial r}+\frac{\partial{v_z}}{\partial z} \right]=-f
 \end{equation}

\begin{equation}
\label{eq14}
  P=P(\rho,T),\quad E=E(\rho,T),\quad
  f=f(\rho,T),
 \end{equation}

Here ${\bf v}=(v_r,\, v_\varphi,\, v_z)$ is the velocity vector, ${\bf B}=(B_r,\, B_\varphi,\, B_z)$ is
the magnetic field vector, ${\bf j}= (j_r,\, j_\varphi,\, j_z)$ is the vector of the electrical current,
 $\rho,\,  P, \, E, \,$ are density, pressure, and internal energy,
respectively, $f$ is the cooling function, due to photons or neutrino.

We consider a long cylinder with a magnetic field directed along its axis.
This cylinder will expand unlimitly under the action of pressure and magnetic forces,
so no confinement will be reached. The limitation of the radius of this cylinder could
be possible in dynamic state, when the whole cylinder undergoes magneto-torsional oscillations.
Such oscillations produce toroidal field, which prevent a radial expansion. There is a
competition between the induced toroidal field, compressing the cylinder in radial direction,
and gas pressure, together with the field along the cylinder axis (poloidal), tending to increase
its radius.During magneto-torsional oscillations there are phases, when either compression or expansion
forces prevail, and, depending on the input parameters, we may expect tree kinds of a behavior of such
cylinder.

1. The oscillation amplitude is low, so the cylinder suffers unlimited expansion (no confinement)

2. The oscillation amplitude is too high, so the pinch action of the toroidal
field destroys the cylinder, and leads to formation of separated blobs.

3. The oscillation amplitude is moderate, so the cylinder survives for an unlimited time,
and its parameters (radius, density, magnetic field etc.) change periodically,
or quasi-periodically in time.

Solution of MHD equations (\ref{eq1})-(\ref{eq14}) could give, in principle,
the answer about a correctness of the above scenario. It is reasonable nevertheless to try
to find a simple approximate way for obtaining a qualitative answer, and to make a rough
estimation of parameters leading to different regimes.

\section{Profiling in axially symmetric MHD equations}

We'll try to simplify the system of equations in such a way, that the resulting system should
contain the most important property of the dynamical competition between different forces, to
check the possibility of the dynamical confinement.We use for this purpose a profiling procedure.
Let us neglect the gravity in the the direction of the cylinder axis ($z$), and approximate
density by a function $\rho(t,z)$, suggesting a uniform density along the radius.
cylinder density. The components of the velocity and magnetic field are approximated as

\begin{equation}
\label{eq15}
  v_r=r\, a(t,z), \quad v_\varphi=r\, \Omega(t,z),\quad v_z=0;
 \end{equation}

\begin{equation}
\label{eq16}
B_r=r\, h_r(t,z), \quad B_\varphi=r\, h_\varphi(t,z), B_z=B_z(t,z).
 \end{equation}
In this case the current components (\ref{eq9})-(\ref{eq11}) are written as

\begin{equation}
\label{eq17}
 j_r=-\frac{cr}{4\pi}\frac{\partial h_\varphi}{\partial z},\quad
 j_\varphi=\frac{cr}{4\pi}\frac{\partial h_r}{\partial z},\quad
 j_z=\frac{ch_\varphi}{2\pi}.
 \end{equation}
After neglecting velocity $v_z$ along the axis, we should omit the corresponding
Euler equation (\ref{eq3}), and the radial pressure gradient is approximated by the linear
function

\begin{equation}
\label{eq18}
\frac{\partial P}{\partial r}= \lambda \frac{P}{R^2}r,
 \end{equation}
where the constant $\lambda \sim 1$ is connected with the equation of state, $P(t,z)$ is
the pressure, $R(t,z)$ is the radius of the cylinder. In the subsequent consideration we
consider an adiabatic case, where the polytropic equation of state
$P=K\,\rho^\gamma$ is considered instead of the energy equation (\ref{eq13}).
Neglecting the $z$ derivatives in the Poisson equation (\ref{eq12}), we obtain
$\varphi_G=\pi G \rho r^2$.
Substituting (\ref{eq15})-(\ref{eq18}) into the original system of the equations, we obtain
for the profiling functions the following equations

\begin{equation}
\label{eq19}
\frac{\partial a}{\partial t}+a^2-\Omega^2=\lambda\frac{P}{\rho R^2}
-2\pi G\rho+\frac{1}{4\pi\rho}\left(B_z\frac{\partial h_r}{\partial z}-2 h_\varphi^2\right),
 \end{equation}

\begin{equation}
\label{eq20}
\frac{\partial \Omega}{\partial t}+2a\Omega=
\frac{1}{4\pi\rho}\left(B_z\frac{\partial h_\varphi}{\partial z}+2 h_r h_\varphi\right),
 \end{equation}

\begin{equation}
\label{eq21}
\frac{\partial h_r}{\partial t}=\frac{\partial (aB_z)}{\partial z},
 \end{equation}

\begin{equation}
\label{eq22}
\frac{\partial h_\varphi}{\partial t}=\frac{\partial (\Omega B_z)}{\partial z}
-2(a h_\varphi-\Omega h_r),
 \end{equation}

\begin{equation}
\label{eq23}
\frac{\partial B_z}{\partial t}=-2 a B_z,
 \end{equation}

\begin{equation}
\label{eq24}
\frac{\partial \rho}{\partial t}=-2 a \rho,
 \end{equation}

\begin{equation}
\label{eq25}
\frac{\partial R}{\partial t}=a R.
 \end{equation}
It follows from (\ref{eq23})-(\ref{eq25}) relations, representing conservation of mass, and
magnetic flux equivalent to freezing condition

\begin{equation}
\label{eq26}
\rho\, R^2=\,C_m(z), \quad B_z\, R^2=\, C_b(z), \quad B_z=\frac{C_b(z)}{C_m(z)}\rho.
 \end{equation}
In our subsequent consideration the arbitrary functions will be taken as constants:
$C_m(z)=C_m, \,\, C_b(z)=C_b.$ The algebraic relations (\ref{eq26}) may be used instead of any two
equations from (\ref{eq23})-(\ref{eq25}).

\section{Equilibrium configuration and linear oscillations}

To check the properties of the approximate system (\ref{eq19})-(\ref{eq25}), we consider
linear oscillations of the equilibrium, infinite, self-gravitating cylinder with uniform
magnetic field and rotation along its axis: $\frac{dB_{z0}}{dz}=0$, $\frac{d\Omega_0}{dz}=0$.
In equilibrium state (index "0") we have

\begin{equation}
\label{eq27}
 a_0=h_{r0}=h_{\varphi 0}=0,\,\, \rho_0={\rm const},\,\, \Omega_0={\rm const},
\,\,\Omega_0^2=2\pi G \rho_0 -\lambda\frac{P_0}{\rho R_0^2}.
 \end{equation}
Linearizing equations (\ref{eq19})-(\ref{eq22}) around the equilibrium state (\ref{eq27}),
we obtain

$$\frac{\partial a}{\partial t}-2\Omega_0\Omega=\lambda\frac{P_0}{\rho_0 R_0^2}
\left(\frac{P}{P_0}-\frac{\rho}{\rho_0}-2\frac{R}{R_0}\right)
-2\pi G\rho+\frac{B_{z0}}{4\pi\rho_0}\frac{\partial h_r}{\partial z},
$$
$$\frac{\partial \Omega}{\partial t}+2a\Omega_0=
\frac{B_{z0}}{4\pi\rho_0}\frac{\partial h_\varphi}{\partial z},
$$
\begin{equation}
\label{eq28}
\frac{\partial h_r}{\partial t}=B_{z0}\frac{\partial (a)}{\partial z},
 \end{equation}
$$\frac{\partial h_\varphi}{\partial t}=\Omega_0\frac{\partial B_z}{\partial z}
+ B_{z0}\frac{\partial \Omega}{\partial z}+2\Omega_0 h_r.
$$
Linearizing of (\ref{eq23})-(\ref{eq25}), using the polytropic equation
of state $P=K\,\rho^\gamma$, and looking for a solution
 in the form $\sim \exp{i(kz-\omega t)}$ we obtain

\begin{equation}
\label{eq29}
\frac{R}{R_0}=i\frac{a}{\omega}, \,\,
\frac{\rho}{\rho_0}=-2\frac{R}{R_0}=-2i\frac{a}{\omega}, \,\,
\frac{P}{P_0}=\gamma\frac{\rho}{\rho_0},\,\,
\frac{B_z}{B_{z0}}=-2\frac{R}{R_0}=-2i\frac{a}{\omega}.
\end{equation}
Here small perturbation values are taken without "0". From the last two equations (\ref{eq28})
we obtain using (\ref{eq29})

\begin{equation}
\label{eq30}
h_r=-B_{z0}\frac{ak}{\omega}, \,\,
h_\varphi=-B_{z0}\frac{\Omega k}{\omega}.
\end{equation}
First two equations (\ref{eq28}) give after using (\ref{eq29}),(\ref{eq30})

$$\left[1-\frac{k^2V_{A0}^2}{\omega^2}-\frac{V_{s0}^2}{R_0^2\omega^2}
\left(1-\frac{1}{\gamma}\right)+2\frac{\Omega_0^2}{\omega^2}\right]a
-2i\frac{\Omega_0}{\omega}\Omega=0
$$
\begin{equation}
\label{eq31}
2i\frac{\Omega_0}{\omega}a+\left(1-\frac{k^2V_{A0}^2}{\omega^2}\right)\Omega=0
\end{equation}
Here we have introduced unperturbed sound $(V_{s0})$, and Alfven $(V_{A0})$ velocities,
determined as

\begin{equation}
\label{eq32}
V_{s0}^2=\gamma\frac{P_0}{\rho_0},\quad V_{A0}^2=\frac{B_{z0}^2}{4\pi\rho_0}.
 \end{equation}
Equations (\ref{eq31}) lead to following dispersion equation

$$
\omega^4-2\left[k^2 V_{A0}^2+\Omega_0^2+\lambda\frac{V_{s0}^2}{ R_0^2}
\left(1-\frac{1}{\gamma}\right)\right]\omega^2
$$
\begin{equation}
\label{eq33}
+k^2 V_{A0}^2\left[k^2 V_{A0}^2
-2\Omega_0^2+2\lambda\frac{V_{s0}^2}{ R_0^2}
\left(1-\frac{1}{\gamma}\right)\right]=0.
 \end{equation}
 Let us consider several particular cases.

 1. Non-rotating, non-magnetized cylinder, $V_{A0}=\Omega_0=0$. It follows from (\ref{eq33})

\begin{equation}
\label{eq34}
\omega^2=2\lambda\frac{V_{s0}^2}{ R_0^2}\left(1-\frac{1}{\gamma}\right).
 \end{equation}
 Equation (\ref{eq34})  describes the only possible mode, remaining after fixing
 $v_z=0$, and uniform density over the radius. It corresponds to homologous mode of
 perturbation, where only radius and density (pressure) are oscillating.
 The frequency (\ref{eq34}) becomes zero at $\gamma=1$ (isotherm). This degeneration is
 connected with the property of self-gravitating cylinder, which equilibrium is neutral
 (takes place at any radius) for the isothermal equation of state. This degeneration
 is equivalent to the well-known case at $\gamma = 4/3$ for a spherical star
 (Chandrasekhar, 1939). While we are not interested in study of homologous oscillations,
 we'll take $\gamma=1$ in all father consideration. We have than from (\ref{eq33}) the
 dispersion equation

\begin{equation}
\label{eq35}
\omega^4-2(k^2 V_{A0}^2+\Omega_0^2)\omega^2
+k^2 V_{A0}^2(k^2 V_{A0}^2-2\Omega_0^2)=0,
\end{equation}
which solution is written as

\begin{equation}
\label{eq36}
\omega^2=k^2 V_{A0}^2+\Omega_0^2\pm \Omega_0
\sqrt{4k^2 V_{A0}^2+\Omega_0^2}.
\end{equation}

2. Rotating non-magnetized cylinder,  $V_{A0}=0,\,\,\Omega_0\ne0$. Here the solution
of dispersion equation
$\omega^2=\Omega_0^2\pm \Omega_0^2$ describes the trivial mode $\omega^2=0$, connected with
pure rotational perturbations, and radial oscillations due to an action of the centrifugal force
$\omega^2=2\Omega_0^2$ (Fridman, Polyachenko, 1985).

3. Non-rotating magnetized cylinder, $V_{A0}\ne 0,\,\,\Omega_0=0$. The solution of the dispersion
equation $\omega^2=k^2 V_{A0}^2$ describes two different types of waves, propagating with Alfven velocity.
To find out the nature of these oscillations note that (\ref{eq31}) at $\gamma=1$, $\Omega_0=0$
reduces to

\begin{equation}
\label{eq37}
\left(1-\frac{k^2V_{A0}^2}{\omega^2}\right)a=0, \quad
\left(1-\frac{k^2V_{A0}^2}{\omega^2}\right)\Omega=0.
\end{equation}
The first type of wave corresponds to perturbation only of the radial velocity,
$a \ne 0$, $\Omega = 0$. It follows that from (\ref{eq3}), that $h_r \ne 0$, $h_\varphi=0$.
This solution corresponds to the Alfven wave along the axis where only radius is perturbed, and
no rotation appears.The second type of wave describes perturbations of angular velocity at constant
radius, $a = 0$, $\Omega \ne 0$, where $h_r = 0$, $h_\varphi\ne 0$. This wave corresponds to a pure
torsional alfven wave along the cylinder axis.To obtain a standing wave ( ${\bf e}_s)$ we need a combination
of two waves running  in opposite directions (${\bf e}_{r\pm})$.

$$
a[{\bf e}_{r+}\cos{(kz-\omega t)}+{\bf e}_{r-}\cos{(kr+\omega t)}]=2a\,{\bf e}_s\,\cos{kr}\, \cos{\omega t},
$$
$$
h_r[{\bf e}_{r+}\cos{(kz-\omega t)}+{\bf e}_{r-}\cos{(kr+\omega t)}]
$$
\begin{equation}
\label{eq38}
=-B_{z0}\frac{ak}{\omega}[{\bf e}_{r+}\cos{(kz-\omega t)}-{\bf e}_{r-}\cos{(kr+\omega t)}]
\end{equation}
$$
=-B_{z0}\frac{ak}{\omega}2\,{\bf e}_s\,\sin{kr}\,\sin{\omega t}=
2h_r\,{\bf e}_s\,\sin{kr}\, \sin{\omega t}.
$$
It is clear that radial velocity $a$ and radial magnetic field $h_r$ are oscillating in opposite phases.
For the torsional wave  we have similar relations:

$$
\Omega[{\bf e}_{r+}\cos{(kz-\omega t)}+{\bf e}_{r-}\cos{(kr+\omega t)}]=2\Omega\,{\bf e}_s\,\cos{kr}\, \cos{\omega t},
$$
$$
h_\varphi[{\bf e}_{r+}\cos{(kz-\omega t)}+{\bf e}_{r-}\cos{(kr+\omega t)}]
$$
\begin{equation}
\label{eq39}
=-B_{z0}\frac{\Omega k}{\omega}[{\bf e}_{r+}\cos{(kz-\omega t)}-{\bf e}_{r-}\cos{(kr+\omega t)}]
\end{equation}
$$
=-B_{z0}\frac{\Omega k}{\omega}2\,{\bf e}_s\,\sin{kr}\, \sin{\omega t}=
2h_\varphi\,{\bf e}_s\,\sin{kr}\, \sin{\omega t}.
$$
The rotational velocity and azimuthal magnetic field are also oscillating in opposite phases.
In reality we may have a mixture of these two degenerate modes where all perturbations are not zero,
what is always takes place in nonlinear case.
We see here that the approximate system of equations describes correctly small
perturbations, connected with radial and torsional modes,
so we expect that these modes will be described correctly in general nonlinear case.

\section{Farther simplification: reducing the problem to ordinary differential equation}

 While in the relativistic jet the self-gravitating force is expected to be much less than
 the magnetic and pressure forces, we neglect gravity in the subsequent consideration. Without gravity
 the equilibrium static state of the cylinder does not exist. We need to solve numerically the
 system of nonlinear  equations (\ref{eq19})-(\ref{eq22}), (\ref{eq25}), (\ref{eq26}) to check the possibility of
 the existence of a cylinder, which radius remains to be finite due to torsional oscillations (dynamic confinement).

We'll try instead to reduce the system to ordinary equations, making additional simplifications.
Let us consider axially symmetric jet moving along $z$-axis with a constant bulk motion velocity,
in the comoving coordinate frame.
In the jet which confinement is reached due to standing magneto-torsional oscillations, there
are points along $z$-axis where rotational velocity always remains zero in this frame. Let us take
$\Omega=0$ in the plane $z=0$. Let us consider standing wave torsional oscillations
with the space period along $z$ axis  equal to $z_0$. Then nodes with
$\Omega=0$ are situated at $z=\pm n \frac{z_0}{2}$, $n=0,1,2,...$.
Let us write the equations, describing the cylinder behavior
in the plane $z=0$, where $\Omega=0$.
All values in this plane we denote by ($\tilde {\rm \phantom{a}}$).
We take also for simplicity $\lambda=1$. We have than equations
in the plane $z=0$ as

\begin{equation}
\label{eq40}
\frac{d{\tilde a}}{d t}+\tilde a^2=\frac{K}{\tilde R^2}
+\frac{C_b}{4\pi C_m}\left(\frac{\partial h_r}{\partial z}\right)_{z=0}
-\frac{\tilde h_\varphi^2\tilde R^2}{2\pi C_m},
\end{equation}
from the equation (\ref{eq19});
\begin{equation}
\label{eq41}
\frac{C_b}{4\pi C_m}\left(\frac{\partial h_\varphi}{\partial z}\right)_{z=0}
+\frac{\tilde h_\varphi \tilde h_r \tilde R^2}{2\pi C_m}=0,
\end{equation}
from the equation (\ref{eq20});
\begin{equation}
\label{eq42}
\frac{d\tilde h_r}{d t}=C_b
\left(\frac{\partial (a/R^2)}{\partial z}\right)_{z=0},
 \end{equation}
from the equation (\ref{eq21});
\begin{equation}
\label{eq43}
\frac{d \tilde h_\varphi}{d t}=
C_b\left(\frac{\partial(\Omega/R^2)}{\partial z}\right)_{z=0}
-2\tilde a \tilde h_\varphi,
\end{equation}
from the equation (\ref{eq22});
\begin{equation}
\label{eq44}
\frac{d \tilde R}{d t}=\tilde a \tilde R.
\end{equation}
from the equation (\ref{eq25}).
The integrals of motion (\ref{eq26}) in the plane $z=0$ are written as
\begin{equation}
\label{eq45}
\tilde \rho\, \tilde R^2=\,C_m, \quad \tilde B_z\, \tilde R^2=\, C_b, \quad \tilde B_z=\frac{C_b}{C_m}\tilde\rho.
\end{equation}
Initial conditions for the system (\ref{eq40}) -  (\ref{eq45}) are

\begin{equation}
\label{eq46}
\tilde R=R_0,\,\, \tilde\rho=\rho_0=\frac{C_m}{ R_0^2}, \,\,
\tilde B_z=\frac{C_b}{ R_0^2},\,\,
\tilde a=\tilde h_r=\tilde h_\varphi=0\,\, {\rm at}\,\, t=0.
 \end{equation}
In (\ref{eq40}) - (\ref{eq45}) we have used relations

\begin{equation}
\label{eq47}
\tilde \rho=\rho_0 \frac{R_0^2}{\tilde R^2}, \,\, \tilde B_z=\rho_0 \frac{C_b}{C_m}\frac{ R_0^2}{\tilde R^2},
 \end{equation}
valid for any time. If the cylinder rotational velocity is antisymmetric
relative to the plane $z=0$, $\Omega=0$, and cylinder  density distribution
is symmetric relative to this plane, then
we have extremum (maximum) of the azimuthal magnetic field $h_\phi$,
with $\left(\frac{\partial h_\varphi}{\partial z}\right)_{z=0}=0$, and zero value of $\tilde h_r=0$, which
reaches an extremum (minimum) in this plane with
$\left(\frac{\partial h_r}{\partial z}\right)_{z=0}=0$.
The product $a\rho$ also reaches an extremum in the plane $z=0$, so that
$\left(\frac{\partial (a/R^2)}{\partial z}\right)_{z=0}=0$.
The term with $z$ derivative in the equation (\ref{eq43}) is not equal to zero, and changes
periodically during the torsional oscillations. We substitute approximately the derivative $d/dz$
by the  ratio $1/z_0$,
where $z_0$ is the space period of the torsional oscillations along $z$ axis. While $\Omega=0$ in the plane
$z=0$, its derivative along $z$ is changing periodically with an amplitude $\Omega_0$,
and frequency $\omega$, which should be
found from the solution of the problem. We approximate therefore

\begin{equation}
\label{eq48}
\left(\frac{\partial(\Omega/R^2)}{\partial z}\right)_{z=0}
=\frac{\Omega_0}{z_0\tilde R^2}\cos{\omega t}.
\end{equation}
Finally, we have from (\ref{eq40}),(\ref{eq43}),(\ref{eq44}) the following approximate system of equations,
describing the non-linear torsional oscillations of the cylinder at given $z_0$ and $\Omega_0$.

$$
\frac{d{\tilde a}}{d t}+\tilde a^2=\frac{K}{\tilde R^2}
-\frac{\tilde h_\varphi^2 \tilde R^2}{2\pi C_m},
 $$
\begin{equation}
\label{eq49}
\frac{d \tilde h_\varphi}{d t}=
C_b \frac{\Omega_0}{z_0\tilde R^2}\cos{\omega t}
-2\tilde a \tilde h_\varphi,
 \end{equation}
$$
\frac{d \tilde R}{d t}=\tilde a \tilde R.
$$
The combination of last two equations gives

\begin{equation}
\label{eq50}
\frac{d (\tilde h_\varphi R^2)}{d t}=
 \frac{C_b\Omega_0}{z_0}\cos{\omega t},
 \end{equation}
with the solution, satisfying initial condition (\ref{eq46}), in the form

\begin{equation}
\label{eq51}
\tilde h_\varphi R^2= \frac{C_b\Omega_0}{z_0\omega}\sin{\omega t}.
 \end{equation}
 With account of (\ref{eq51}) the first and third equations in (\ref{eq49}) give the
 equation

\begin{equation}
\label{eq52}
\tilde R\frac{d (\tilde a R)}{d t}=
 K-\left(\frac{C_b\Omega_0}{z_0\omega}\right)^2\frac{\sin^2{\omega t}}{2\pi C_m},
 \end{equation}
Two differential equations, (\ref{eq52}) and the third equation (\ref{eq49}) determine
the behavior of the cylinder during magneto-torsional oscillations. Solutions
where the radius does not go to infinity with time, determine a dynamically confined cylinder.
Formation of blobs occurs, when the radius tends to zero. Long dynamically confined cylinder
exist when its radius is changing with time between two finite values.

\section{Numerical solution}

Introduce non-dimensional variables

\begin{equation}
\label{eq53}
\tau=\omega t, \,\, y=\frac{\tilde R}{R_0},\,\, z=\frac{a \tilde R}{a_0 R_0},\,\,
 a_0=\frac{K}{\omega R_0^2}=\omega,\,\,R_0=\frac{\sqrt K}{\omega},
 \end{equation}
in which differential equations have a form

\begin{equation}
\label{eq54}
\frac{dy}{d\tau}=z,\quad
\frac{d z}{d\tau}=\frac{1}{y}(1-D\sin^2 \tau);\,\,\,\,y(0)=1,\,\, z=0\,\,{\rm at}\,\, \tau=0.
 \end{equation}
Therefore, the problem is reduced to a system (\ref{eq54}) with only two non-dimensional parameters
$D=\frac{1}{2\pi K C_m}\left(\frac{C_b\Omega_0}{z_0\omega}\right)^2$, and $y(0)$, and
the second one is taken equal to unity in farther consideration.
All qualitatively different solutions are reproduced inside this restricted set of parameters.
Solution of this nonlinear system
changes qualitatively with changing of the parameter $D$.

The solution of this system was obtained numerically for $D$= 2, 2.1, 2.11, 2.15, 2.2, 2.25, 2.28,
2.4, 2.5, 2.6, 2.9, 3.0. Roughly the solutions may be divided into 3 groups.

1. At $D \le 2$ there is no confinement, and radius grows to infinity after several low-amplitude
oscillations (see Fig.1).

2. With growing of $D$ the amplitude of oscillations increase, and
at $D=2.1$ radius is not growing to infinity, but is oscillating
around some average value, forming rather complicated curve (Figs.
2-4).

3. At $D$= 2.28 and larger the radius finally goes to zero with
time, but with different behavior, depending on $D$. At $D$ between
2.28 and 2.9 the dependence of the radius $y$ with time may be very
complicated, consisting of low-amplitude and large-amplitude
oscillations, which finally lead to zero. The time at which radius
becomes zero depends on $D$ in  rather peculiar way, and may happen
at $\tau \le 100$, like at $D$=2.4, 2.6 (Figs. 7,10); or goes trough
very large radius, and returned back to zero value at very large
time $\tau \sim 10^7$ at $D$=2.5 (Fig. 9). Starting from $D=3$ and
larger the solution becomes very simple, and radius goes to zero at
$\tau<2.5$ (Fig. 13), before the right side of the second equation
(\ref{eq54}) returned to the positive value. The results of
numerical solution are represented in Figs. 1-13.

% Fig.2
\begin{figure}[p]
\centerline
{
\psfig{figure=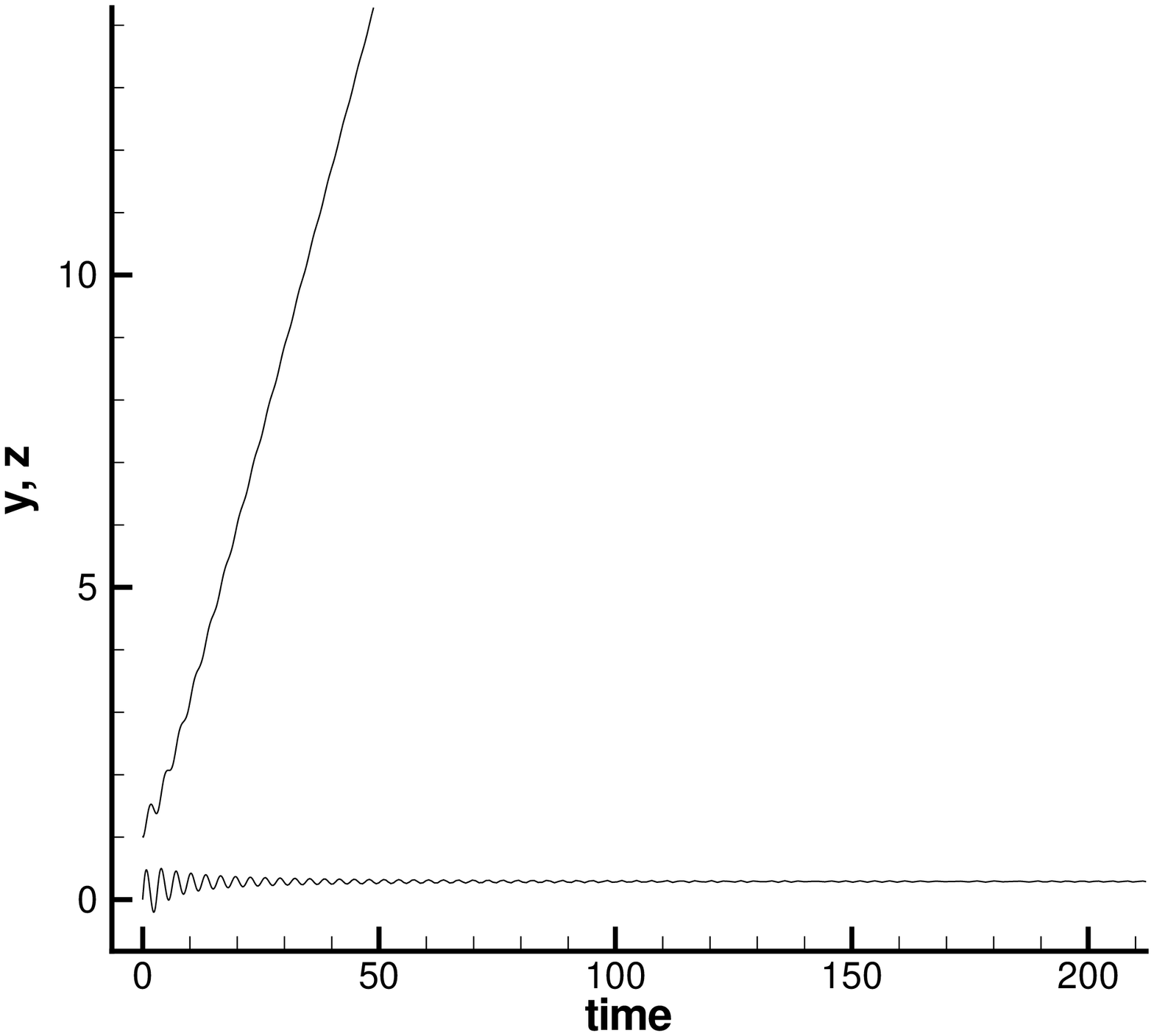,width=10cm,angle=-90}
}
\caption{Time dependence of non-dimensional radius $y$ (upper curve), and non-dimensional velocity
$z$ (lower curve), for $D=2.0$.}
\label{fig2}
\end{figure}

% Fig.3
\begin{figure}[p]
\centerline
{
\psfig{figure=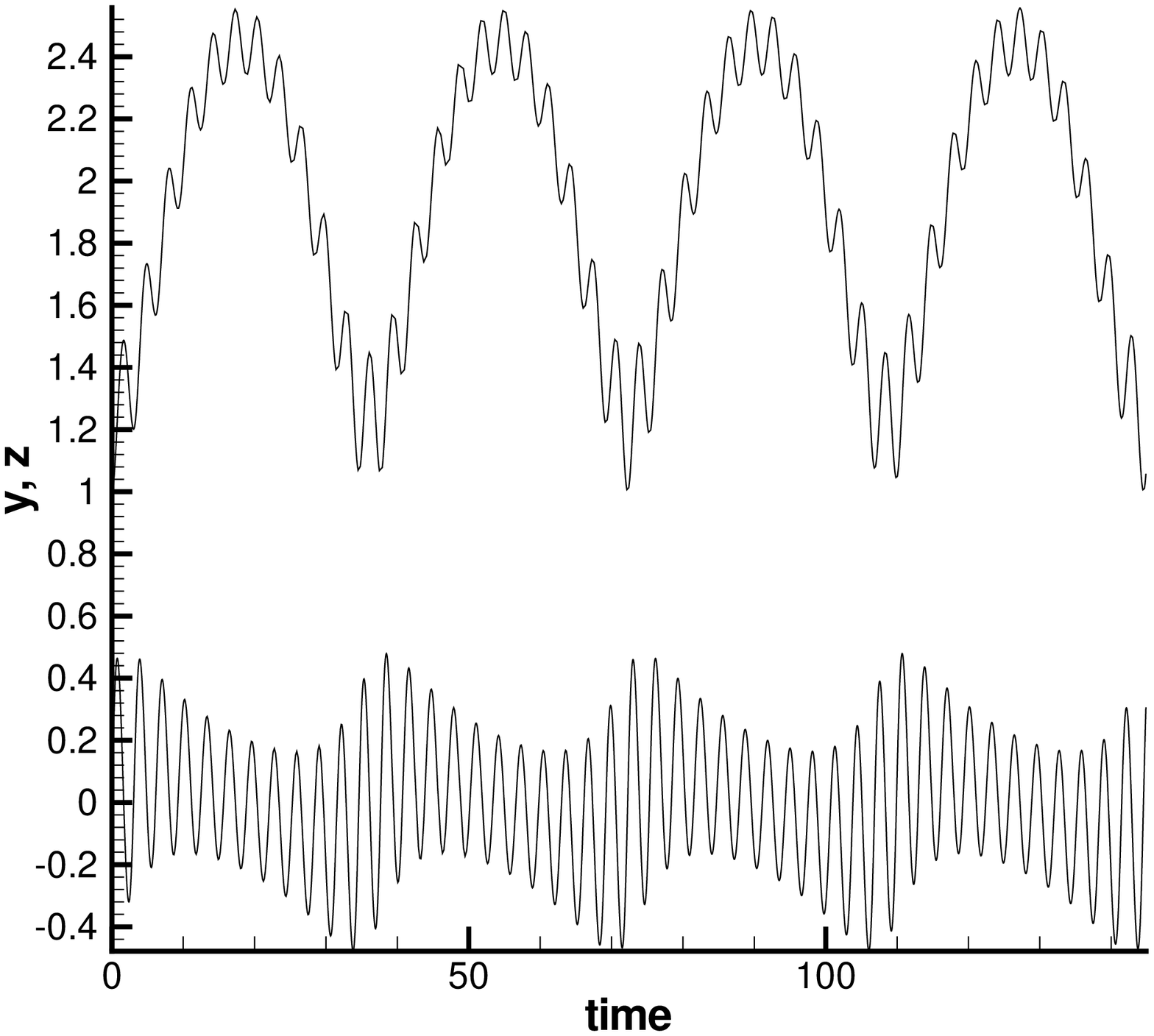,width=10cm,angle=-90}
}
\caption{Time dependence of non-dimensional radius $y$ (upper curve), and non-dimensional velocity
$z$ (lower curve), for $D=2.1$.}
\label{fig3}
\end{figure}

% Fig.3a
\begin{figure}[p]
\centerline
{
\psfig{figure=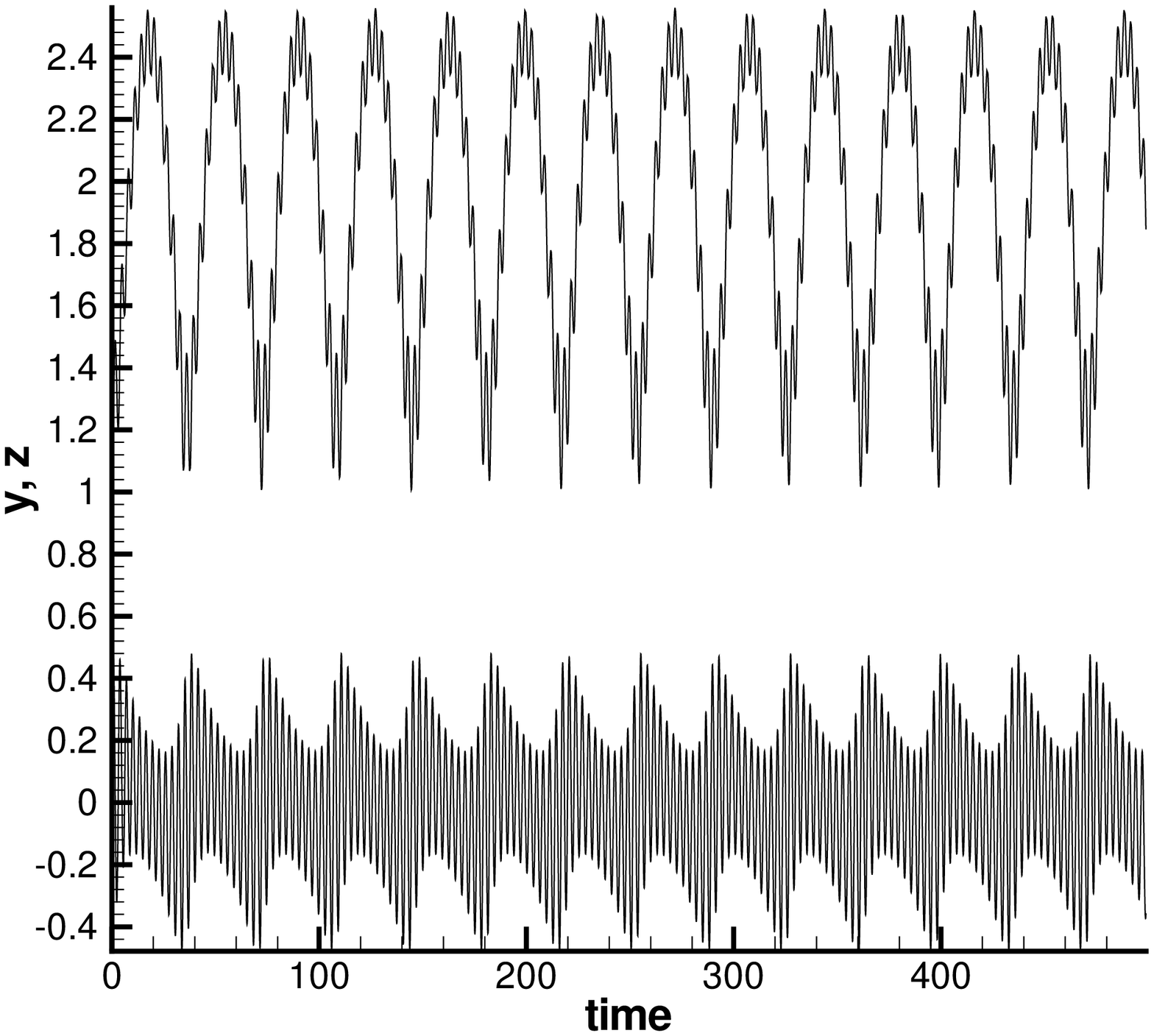,width=10cm,angle=-90}
}
\caption{Time dependence of non-dimensional radius $y$ (upper curve), and non-dimensional velocity
$z$ (lower curve), for $D=2.1$ during a long time period.}
\label{fig3a}
\end{figure}

%Fig.7a
\begin{figure}[p]
\centerline{
\psfig{figure=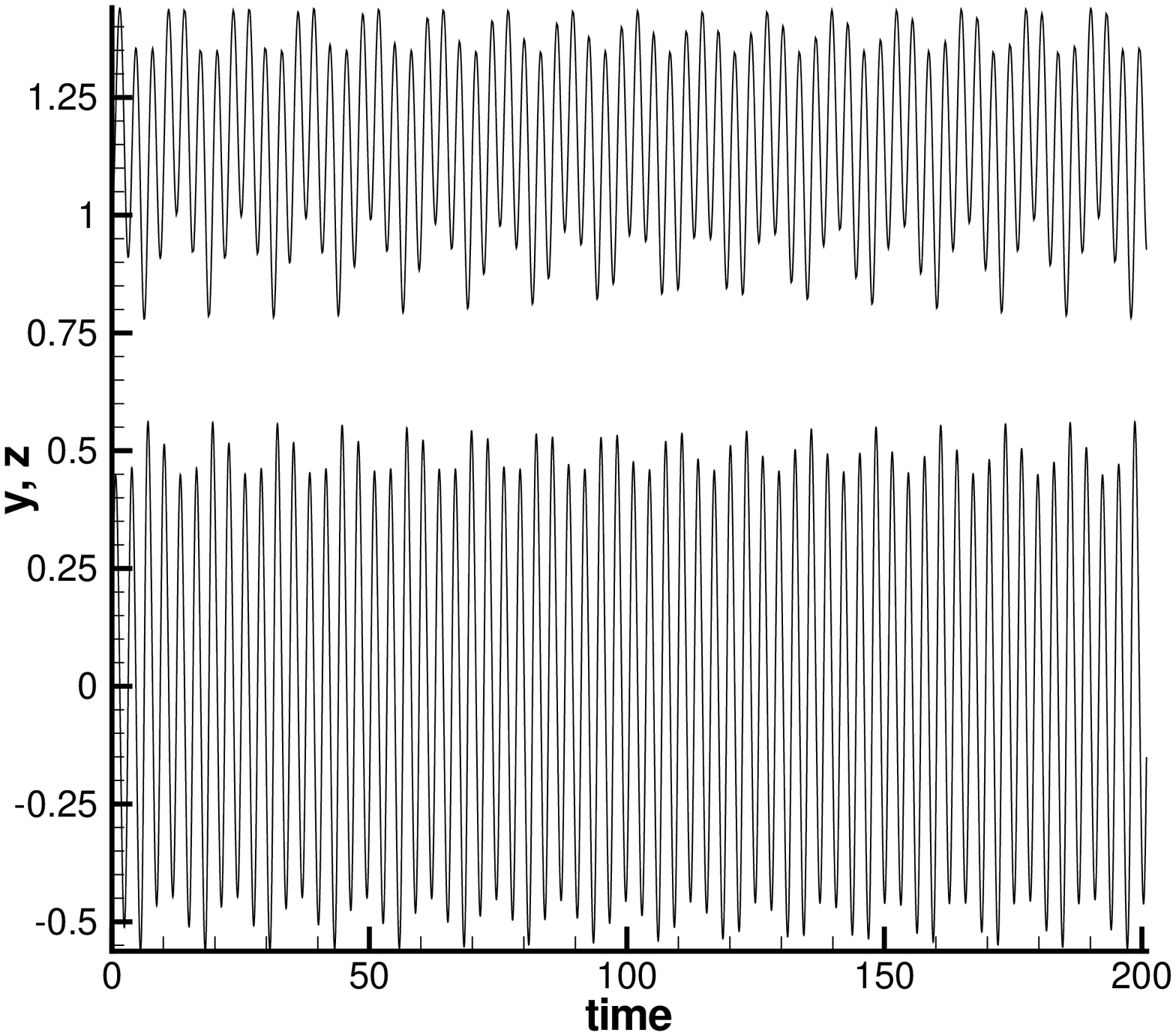,width=10cm,angle=-90}
}
\caption{Time dependence of non-dimensional radius $y$ (upper curve), and non-dimensional velocity
$z$ (lower curve), for $D=2.25$, during a long time period.}
\label{fig7a}
\end{figure}

%Fig.8
\begin{figure}[p]
\centerline{ \psfig{figure=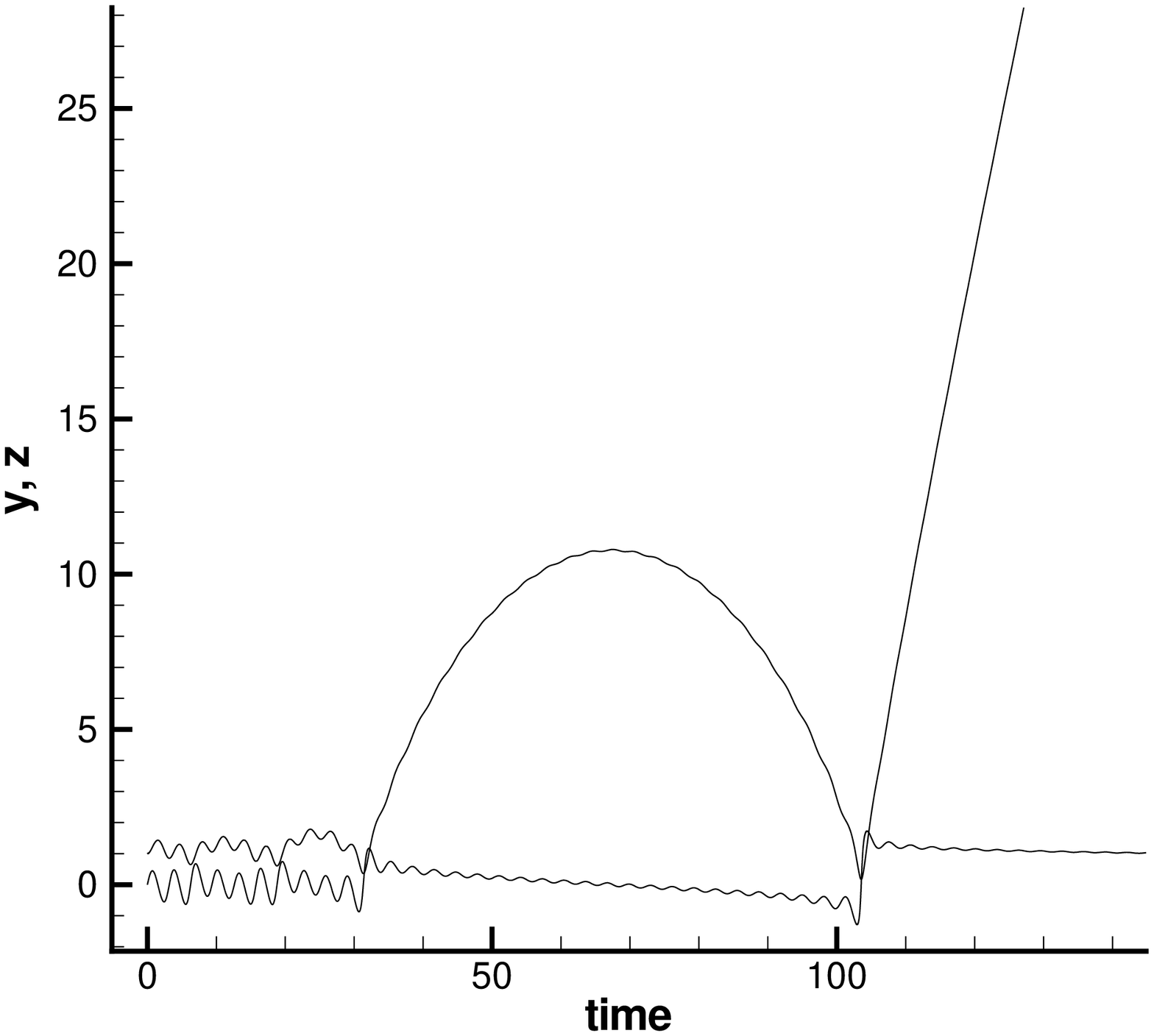,width=10cm,angle=-90} }
\caption{Time dependence of non-dimensional radius $y$ (upper curve), and non-dimensional velocity
$z$ (lower curve), for $D=2.28$.}
\label{fig8}
\end{figure}

%Fig.8a
\begin{figure}[p]
\centerline{ \psfig{figure=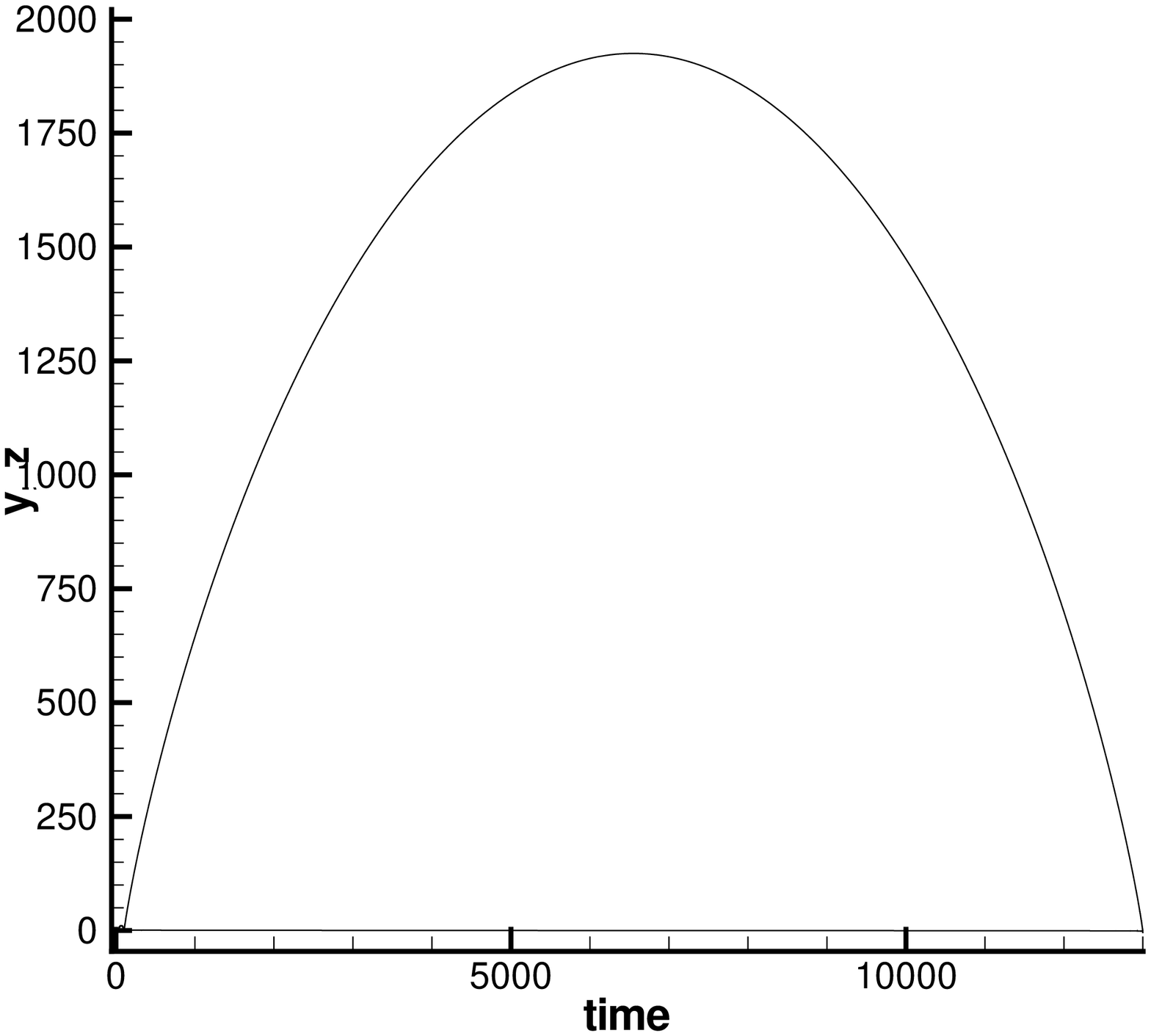,width=10cm,angle=-90} }
\caption{Time dependence of non-dimensional radius $y$ (upper curve), and non-dimensional velocity
$z$ (lower curve), for $D=2.28$, during a long time period.}
\label{fig8a}
\end{figure}

%Fig.10
\begin{figure}[p]
\centerline{ \psfig{figure=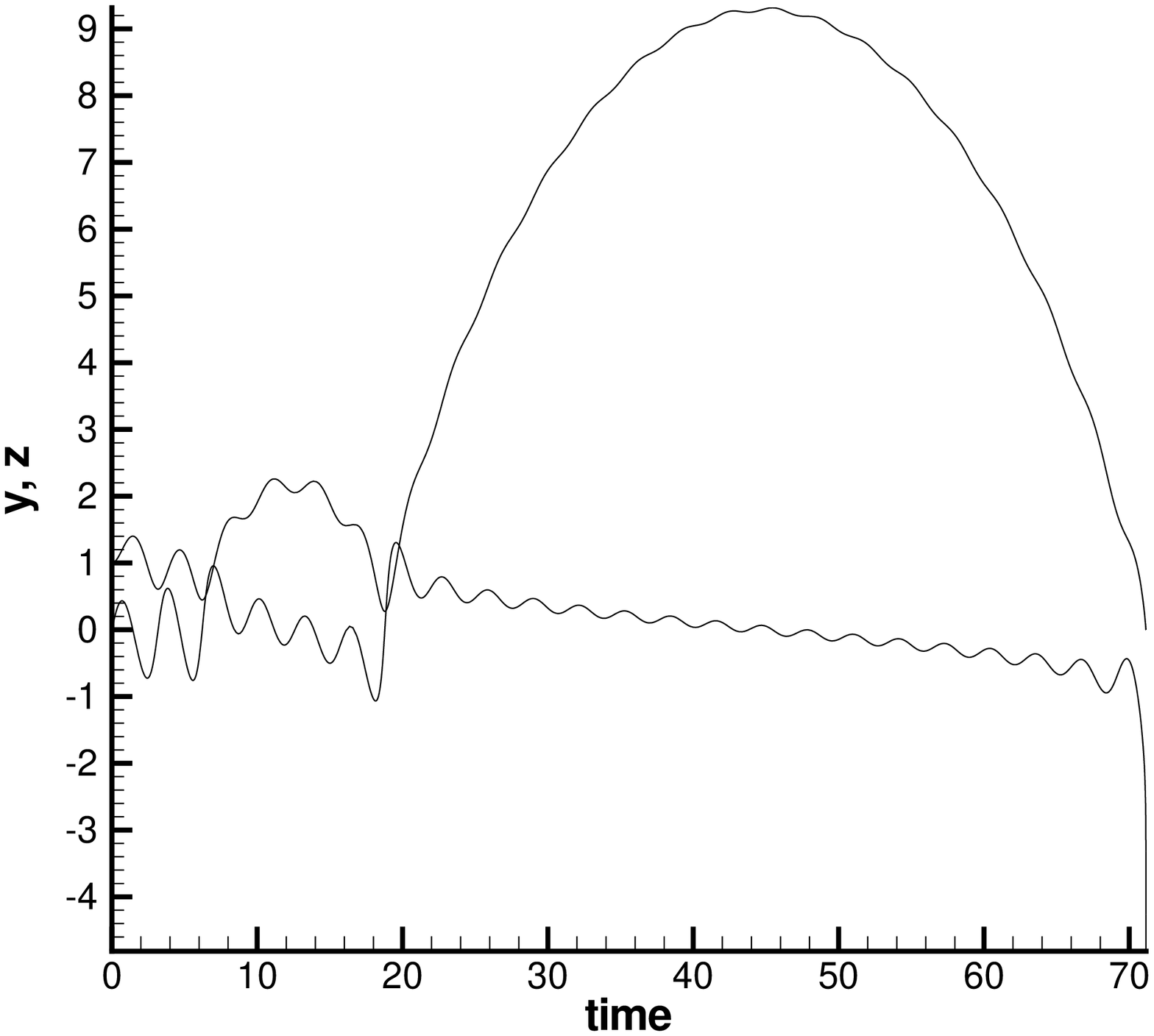,width=10cm,angle=-90} }
\caption{Time dependence of non-dimensional radius $y$ (upper curve), and non-dimensional velocity
$z$ (lower curve), for $D=2.4$.}
\label{fig10}
\end{figure}

\section{Discussion}

Let us consider, a jet which has an equation of state in the form
$P=K\rho=v_s^2 \rho$, $v_s^2 \le c^2/3$, $v_s$ is the sound speed in
the matter. For ultrarelativistic pair-plasma we have $P=c^2/3$. The
non-dimensional parameter $D$, as a function of the characteristic
radius $R_0$, periodic length $z_0$ along $z$ axis, initial density
$\rho_0$, and magnetic field $B_{z0}$, $\Omega_0$ and $\omega$ is
written in the form

\begin{equation}
\label{eq55} D=\frac{1}{2\pi\rho_0}\frac{B_{z0}^2 R_0^2
\Omega_0^2}{z_0^2 \omega^2 v_s^2}.
 \end{equation}
The amplitude of oscillations $\Omega$, and $\omega$ should be found
from the solution of the nonlinear system (\ref{eq19})-(\ref{eq25}),
together with determination of the interval of values of $D$ at
which confinement happens. In the approximate system (\ref{eq54})
only one parameter $D$ characterizes different regimes, what for
given values of  $R_0$, $z_0$, $\rho_0$, and $B_{z0}$ determines a
function $\Omega_0(D,\omega)$ for the collimated jet. To find
approximately a self-consistent model with $\Omega_0,\omega(D)$ we
may use the solution of linearized equations
(\ref{eq19})-(\ref{eq25}) with $\omega=k V_A$ from (\ref{eq37}). The
frequency of non-linear oscillations is smaller, and we may write

\begin{equation}
\label{eq56} \omega=\alpha_n\,k\, V_A,\,\, \alpha_n < 1, \,\,
k=\frac{2\pi}{z_0},
 \end{equation}
so that

\begin{equation}
\label{eq57} \omega^2=\alpha_n^2\,k^2\,V_A^2\,=\,\alpha_n^2
\frac{\pi B_{z0}^2}{\rho_0 z_0^2}.
 \end{equation}
Using it in (\ref{eq55}), with account of (\ref{eq53}), we obtain

\begin{equation}
\label{eq58} \Omega^2 R_0^2=2\pi^2 D \alpha_n^2 v_s^2 <c^2,\,\,
R_0^2=\frac{K}{\omega^2}=z_0^2\frac{\rho_0 v_s^2}{\alpha_n^2 \pi B_{z0}^2}.
 \end{equation}
On the edge of the cylinder the rotational velocity cannot exceed the light velocity,
so the solution with initial conditions in (\ref{eq54}), corresponding to  $y_0=1$, has
a physical sense only at $v_s^2 < \frac{c^2}{2\pi^2 D \alpha_n^2} \approx
\frac{c^2}{40 \alpha_n^2}$. Taking $\alpha_n^2=0.1$ for a strongly non-linear oscillations
we obtain a very moderate restriction $v_{s0}^2 < \frac{c^2}{4}$. While in the
intermediate collimation regime the outer tangential velocity is not changing significantly,
this restriction would be enough also for the whole period of the time. To have the sound velocity
not exceeding $c/2$, the jet should contain baryons, which density $\rho_0$
cannot be very small, and its
input in the total density in the jet should be larger than about
30\%.

The confinement by torsional oscillations starts at $D=2.1$, and at
$D\ge 2.28$ the jet is divided into separate blobs according to
Figs.9-17.
So, the confinement by magneto-torsional oscillations can be realized in the
physically available situation, what is the main conclusion of this paper.

\medskip

{\bf Acknowledgement} This work was partially supported by RFBR
grants 05-02-17697, 06-02-91157, 06-02-90864, and President grant
for a support of leading scientific schools 10181.2006.2. Author
started this work during a research programm
 "Physics of Astrophysical Outflows and Accretion Disks" in
KITP UCSB, 2005, and was supported in part by the National Science Foundation under
Grant No. PHY99-0794.

\medskip

%\newpage

\newpage

%Fig.11
\begin{figure}[p]
\centerline{ \psfig{figure=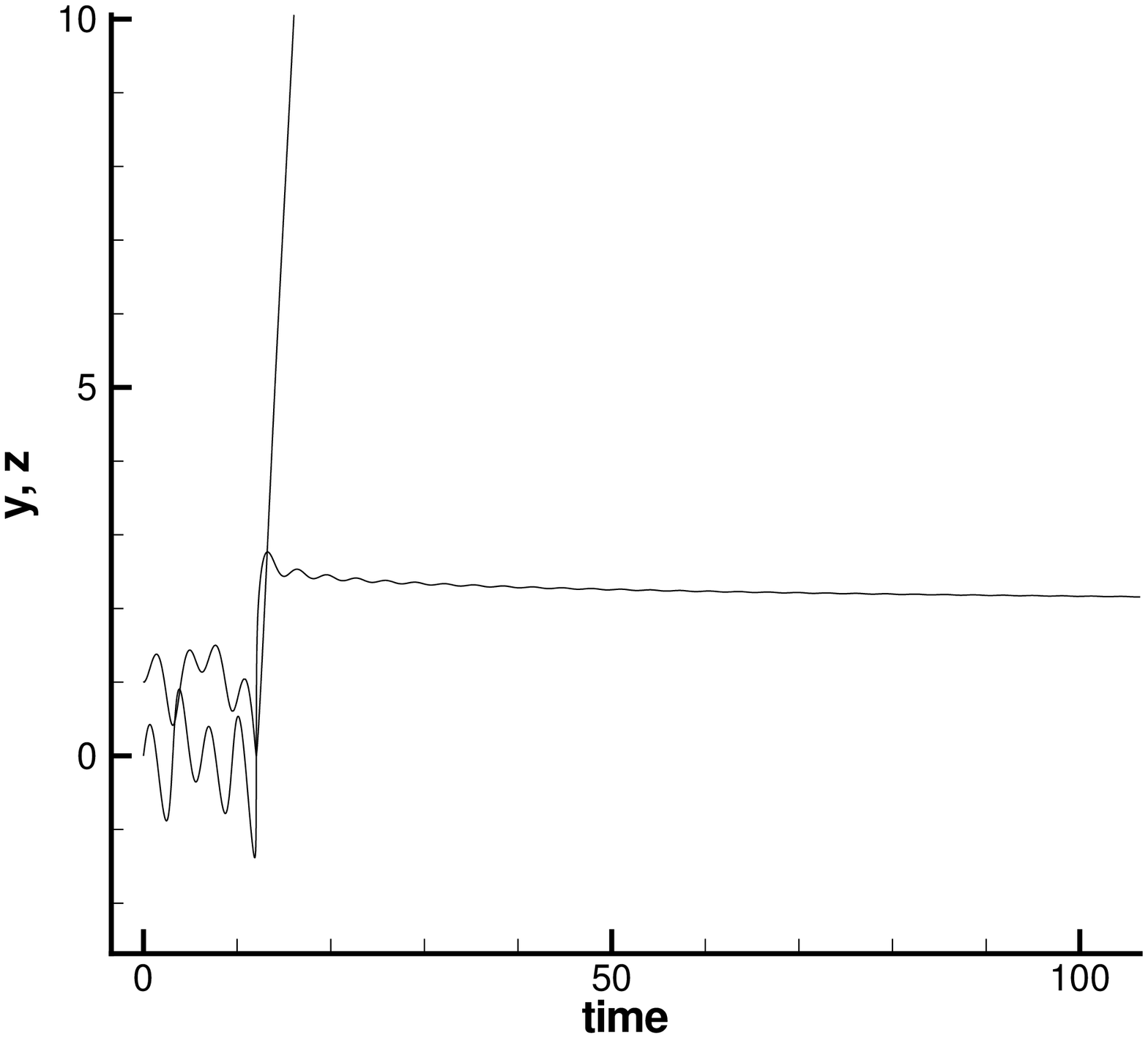,width=10cm,angle=-90} }
\caption{Time dependence of non-dimensional radius $y$ (upper curve), and non-dimensional velocity
$z$ (lower curve), for $D=2.5$.}
\label{fig11}
\end{figure}

%Fig.11a
\begin{figure}[p]
\centerline{ \psfig{figure=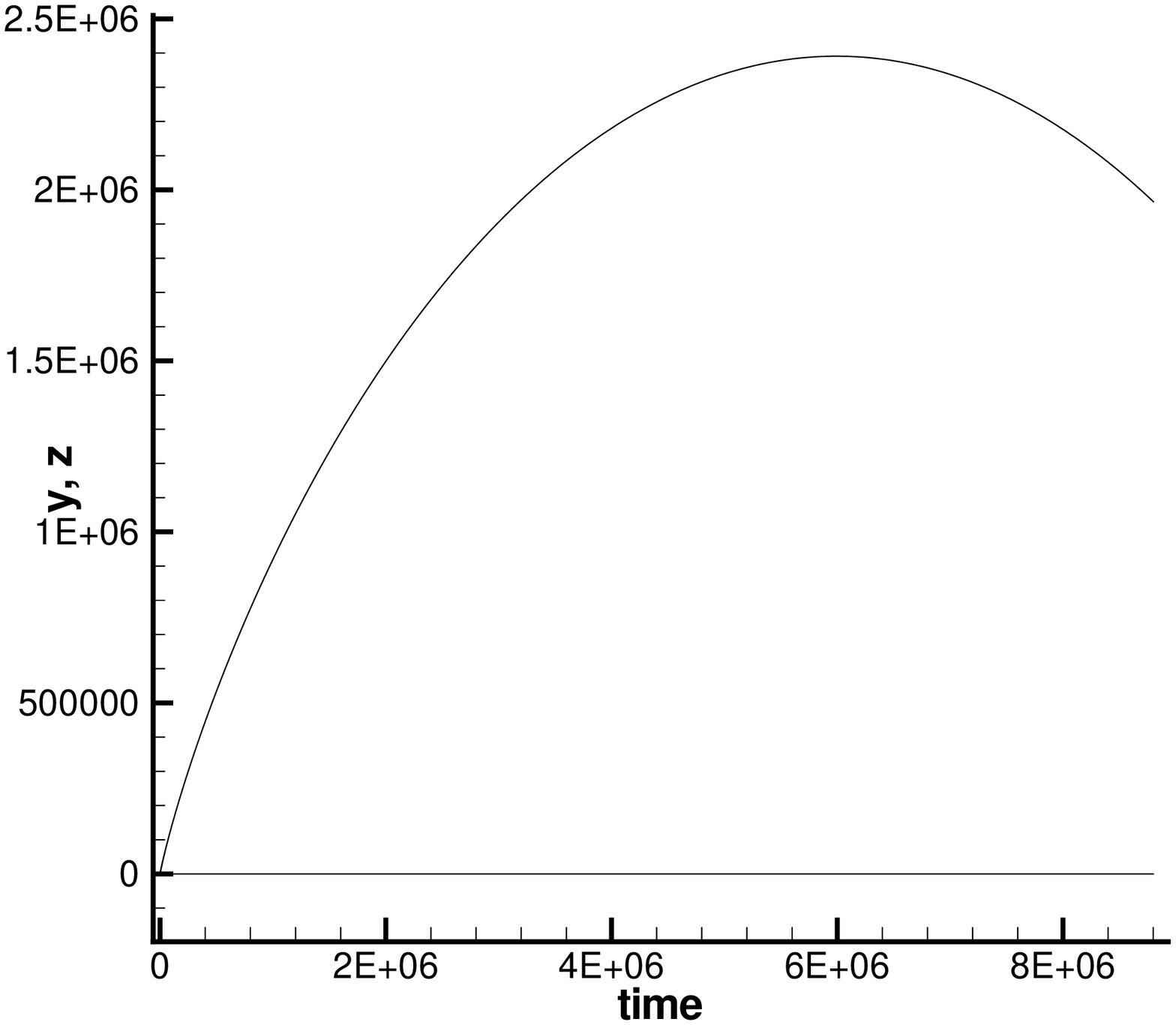,width=10cm,angle=-90} }
\caption{Time dependence of non-dimensional radius $y$ (upper curve), and non-dimensional velocity
$z$ (lower curve), for $D=2.5$, during a long time period.}
\label{fig11a}
\end{figure}

%Fig.13
\begin{figure}[p]
\centerline{ \psfig{figure=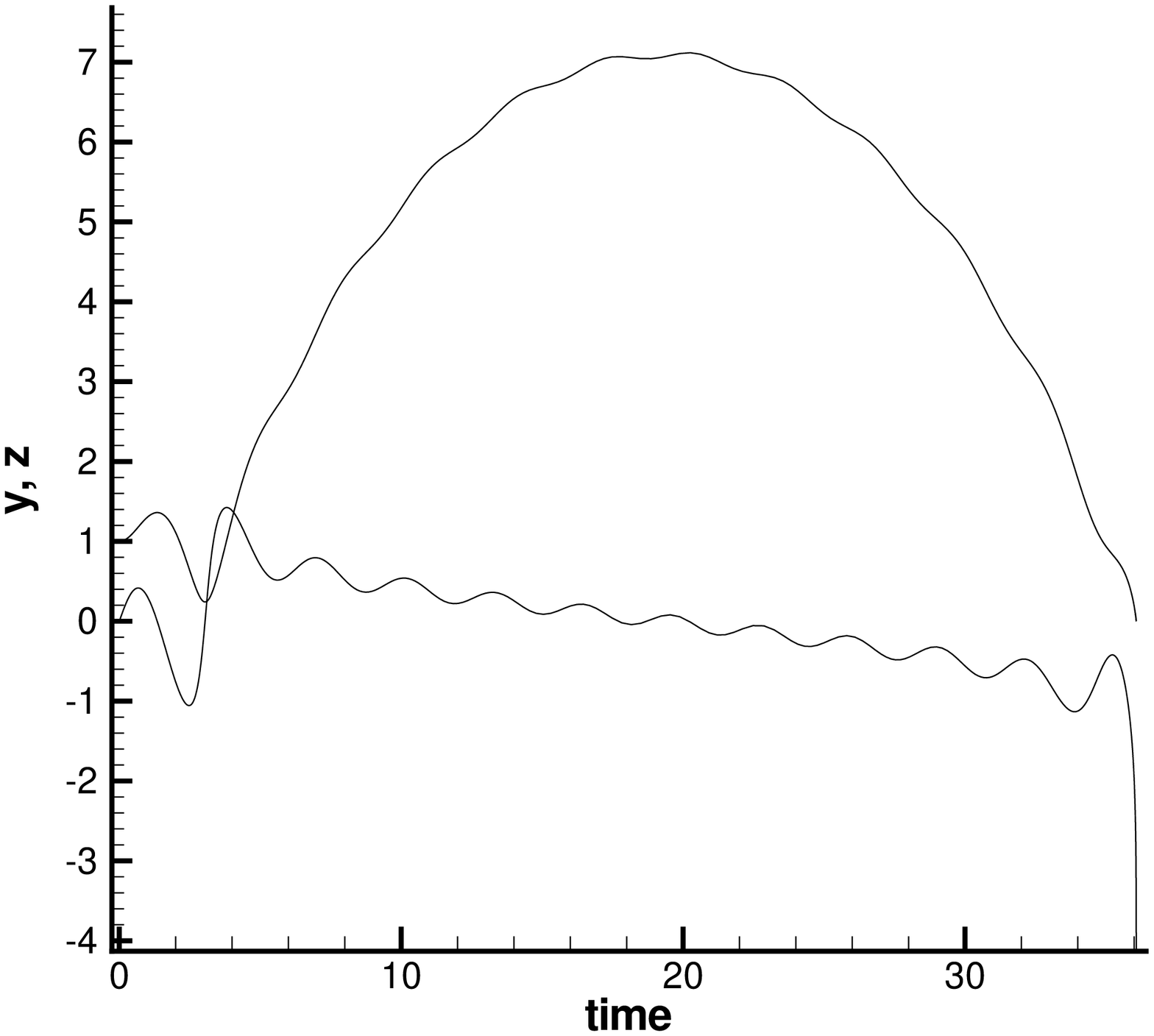,width=10cm,angle=-90} }
\caption{Time dependence of non-dimensional radius $y$ (upper curve), and non-dimensional velocity
$z$ (lower curve), for $D=2.6$.}
\label{fig13}
\end{figure}

%Fig.16
\begin{figure}[p]
\centerline{ \psfig{figure=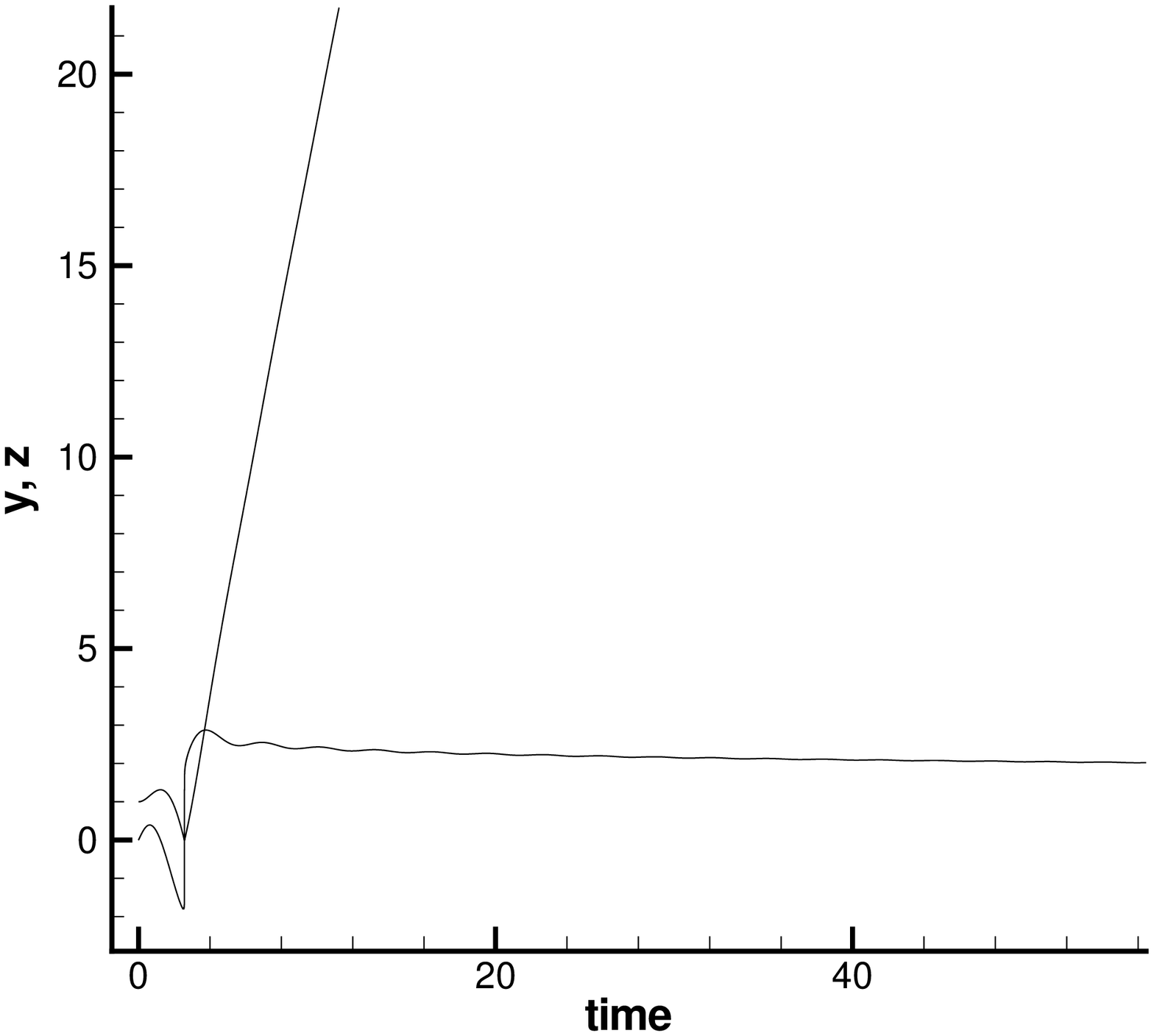,width=10cm,angle=-90} }
\caption{Time dependence of non-dimensional radius $y$ (upper curve), and non-dimensional velocity
$z$ (lower curve), for $D=2.9$.}
\label{fig16}
\end{figure}

%Fig.16a
\begin{figure}[p]
\centerline{ \psfig{figure=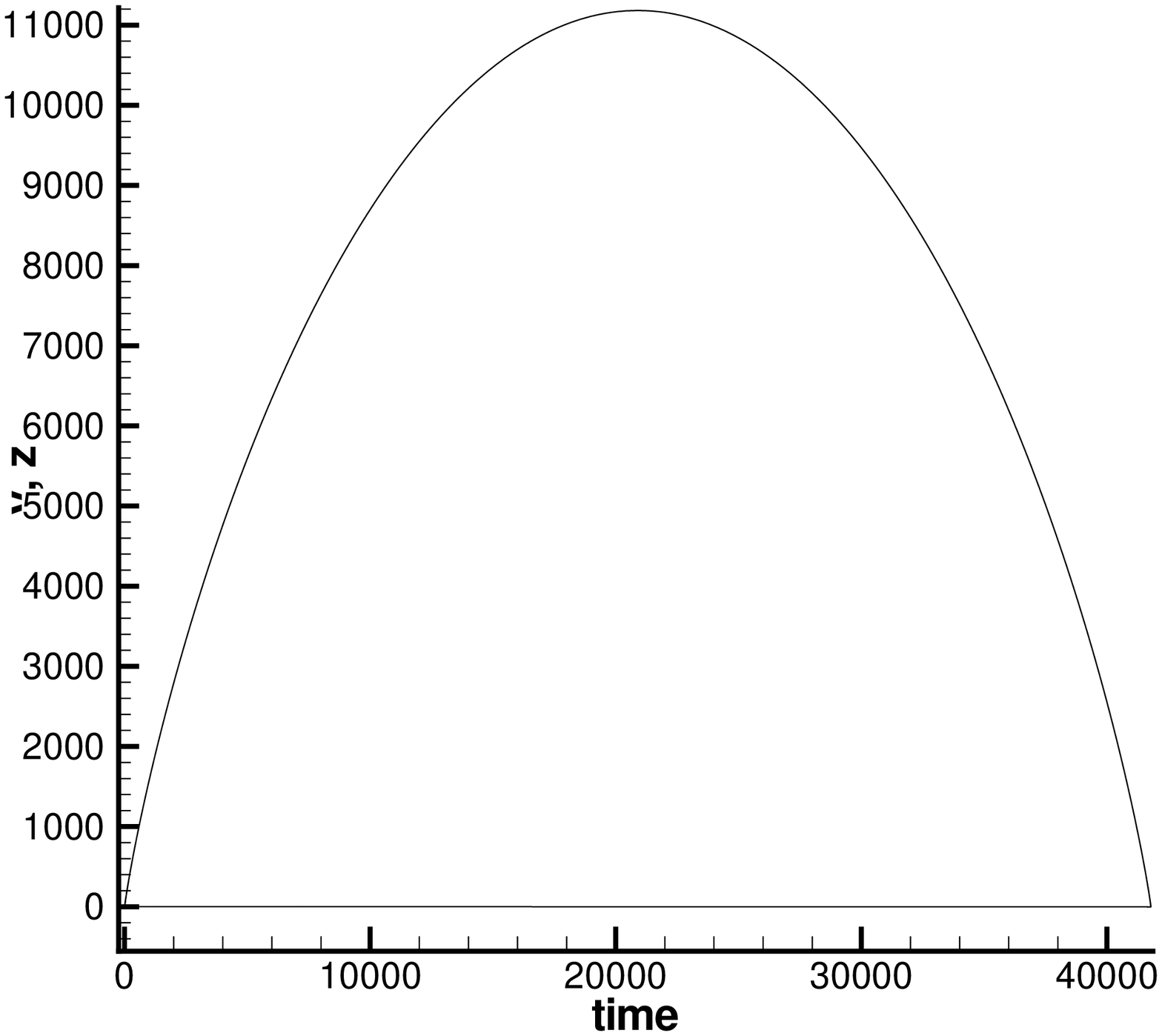,width=10cm,angle=-90} }
\caption{Time dependence of non-dimensional radius $y$ (upper curve), and non-dimensional velocity
$z$ (lower curve), for $D=2.9$, during a long time period.}
\label{fig16a}
\end{figure}

%Fig.17
\begin{figure}[p]
\centerline{ \psfig{figure=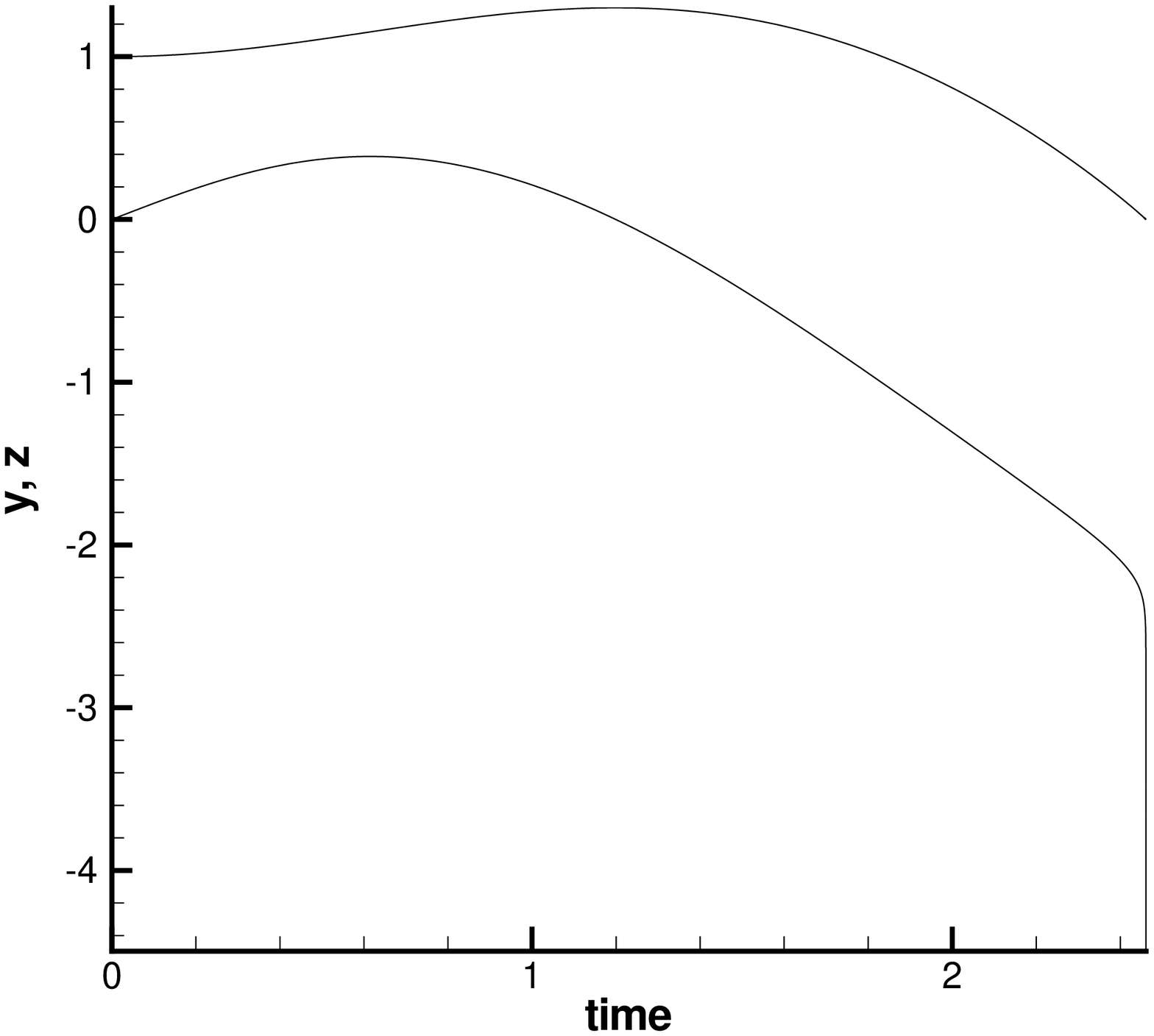,width=10cm,angle=-90} }
\caption{Time dependence of non-dimensional radius $y$ (upper curve), and non-dimensional velocity
$z$ (lower curve), for $D=3.0$.}
\label{fig17}
\end{figure}

\end{document}